\documentclass[final,5p,twocolumn]{elsarticle}

\usepackage{graphicx}
\usepackage{amsmath,amssymb}
\usepackage{booktabs}
\usepackage{float}
\usepackage{subcaption}
\usepackage{placeins}
\usepackage{cuted}
\usepackage{algorithm}
\usepackage{algpseudocode}
\usepackage[hidelinks]{hyperref}

\begin{document}

\begin{frontmatter}

\title{AI-Integrated Learning Management System for Middle School: A Longitudinal Study of Learning Outcomes Through High School and Beyond}

\author[inst1]{Misan Paul Etchie\corref{cor1}}
\ead{mpe45@nau.edu}
\cortext[cor1]{Corresponding author}

\author[inst2]{Olutosin Taiwo}
\ead{Olutosin.Taiwo@nau.edu} 

\affiliation[inst1]{organization={Northern Arizona University},
  city={Flagstaff}, state={AZ}, country={USA}}

\affiliation[inst2]{organization={Northern Arizona University},
  city={Flagstaff}, state={AZ}, country={USA}}

\begin{abstract}
Middle school is a key window for building core academic skills and the learning routines students carry into later grades, yet many students still fall behind because help is often limited and comes too late, after they have already been stuck for a while. Learning Management Systems (LMSs) are now standard infrastructure for distributing materials, collecting work, assessing students' tasks, and recording grades, but in most deployments they still behave more like workflow tools than instructional supports. The result is the usual bottleneck: students keep practicing through confusion, teachers triage questions, and feedback that could have corrected the misunderstanding arrives after the misconception has already hardened. To address this gap, we propose an AI-integrated LMS for middle school instruction, paired with a longitudinal study design to test whether sustained, bounded AI support changes outcomes through high school and into post-high school pathways. The proposed platform adds policy-gated AI assistance to everyday coursework, delivering formative feedback and hinting, recommending spaced review and adaptive practice based on mastery, and providing teacher-facing dashboards that summarize misconception patterns and flag sustained struggle. Because the platform is intended for minors, the design is privacy-first, using data minimization, role-based access control, age-appropriate response constraints, and auditable logs of AI interactions. Beyond short-term performance, the evaluation plan links fine-grained learning traces (attempts, revisions, help-seeking, and pacing) to institutional outcomes where feasible, so we can separate tool adoption effects from longer-run changes in learning trajectories.
\end{abstract}

\begin{keyword}
AI in education \sep Learning management systems \sep Longitudinal study \sep Middle school
\end{keyword}

\end{frontmatter}

\section{Introduction}

During foundational learning, students are expected to make steady learning progress, building academic confidence along the way, regardless of age, background, location, or prior achievement. In practice, those goals are often limited by uneven access to individualized support, variation in instructional resources, differences in home learning environments, and disruptions that reduce instructional time. The “two sigma” tutoring result is a classic illustration of how much individualized support can matter, whether this be one-on-one tutoring or small group sessions. While these are rarely feasible at scale, they produce substantially larger learning gains than conventional classroom instruction \cite{Bloom1984TwoSigma}. At the other end of the spectrum, the COVID-19 pandemic has been able to illustrate how system-level disruptions hinder learning by reducing instructional time, with evidence that achievement gaps widened for students who lacked strong learning support at home \cite{Hammerstein2021COVIDReview}.

These constraints are particularly consequential in middle school, when students consolidate and build foundational skills in reading, writing, and mathematics and develop learning habits such as persistence, planning, and effective help-seeking \cite{EcclesRoeser2011Adolescence,YeagerWalton2011}. The cruciality of this stage cannot be overstated; when students struggle during this stage, misconceptions can solidify, and motivation can decline, making later coursework harder to access. Consistent with this, learning gaps in the middle grades are linked to downstream academic risk, including greater likelihood of course failure and lower odds of on-time progression later in life \cite{Balfanz2007Dropout}.

In recent times, learning management systems (LMSs) have evolved into standard infrastructure for organizing instruction, distributing materials, collecting submissions, and reporting grades with wide-scale integration across varying levels of education \cite{DahlstromBrooks2014ECAR}. This mass adoption has improved coordination between teachers and students and has made learning artifacts and performance records more accessible to many. Yet, despite these operational benefits, conventional LMS platforms typically provide limited instructional support at the moment it is most needed: when a student is confused, stuck, or repeatedly making the same error. In practice, the learning loop can remain slow: a student submits work, waits for feedback, and may continue practicing misconceptions in the meantime. Timely formative feedback is a key driver of learning and dismantling misconceptions right as they occur. Delays simply reduce its effectiveness \cite{HattieTimperley2007}.

Recent advances in artificial intelligence (AI), learning analytics, and educational data infrastructure make it increasingly realistic to embed timely and iterative responsive support directly inside everyday LMS workflows. In practical terms, AI can help shorten the feedback loop by generating formative feedback, offering structured hints, recommending targeted practice, and surfacing class-level misconception patterns that teachers can act on \cite{VanLehn2011MetaAnalysisTutoring,Koedinger2013}. These supports can be made more context-sensitive by conditioning on what a student has already mastered and when they last practiced a skill, which connects naturally to spacing and mastery-based scheduling \cite{PavlikAnderson2008Spacing}. In parallel, help-seeking traces (when students ask for hints, how often, and after what kinds of errors) provide useful signals for detecting struggle and deciding when to escalate support or prompt teacher intervention, given established links between effective help-seeking and academic outcomes \cite{Karabenick2003HelpSeeking}.

At the same time, using AI with minors raises serious concerns and non-negotiable requirements around privacy, safety, transparency, and equity. Educational AI systems need to built from the ground up to minimize exposure of student data, enable accountability, and actively mitigate bias, rather than treating these as afterthoughts \cite{Holmes2021EthicsAIED}. They also need to be designed to support learning rather than replace it: answer-giving behaviors can short-circuit productive struggle, so assistance should emphasize iteration, scaffolding, hints, and guidance that keeps students doing the thinking \cite{Aleven2003HelpSeeking,VanLehn2011MetaAnalysisTutoring}. For these reasons, AI support should preserve teacher agency and produce auditable records of AI–student interactions so that schools can monitor behavior, investigate failures, and evaluate impacts with appropriate oversight \cite{Holmes2021EthicsAIED}. Their built-in answer-giving avoidant behaviors will avoid undermining learning and instead support productive struggle and scaffolded guidance \cite{Aleven2003HelpSeeking,VanLehn2011MetaAnalysisTutoring}. Accordingly, AI support should preserve teacher agency and provide auditable logs of AI--student interactions for oversight and evaluation \cite{Holmes2021EthicsAIED}.

Furthermore, the field needs evidence that extends beyond short-term engagement metrics or immediate test score improvements, examining whether early AI-supported learning produces durable benefits across key long-term educational transitions. Many educational interventions show fade-out over time, motivating the need for longitudinal designs that connect learning experiences to later academic outcomes \cite{Bailey2020Fadeout}.

In this paper, we propose an AI-integrated learning management system for middle school instruction and a longitudinal study framework to assess its effects through high school and into post--high school pathways. The proposed platform integrates existing LMS functionality with policy-gated AI assistance that provides bounded formative feedback and hinting, adaptive sequencing and spaced review recommendations, and study planning and reflection prompts that support self-regulation. A teacher-facing analytics dashboard summarizes misconception trends, highlights students experiencing sustained difficulty, and supports timely instructional interventions. The system is designed to operate as a complement to classroom instruction rather than a replacement, and it implements privacy-first data collection, role-based access control, and auditability of AI outputs to enable safe deployment and rigorous evaluation in real school environments.

The technical and research contributions of this paper are as follows:
\begin{itemize}
    \item A proposal for an AI-integrated LMS that embeds bounded, curriculum-aligned AI support into everyday middle school coursework to enable timely feedback, adaptive practice, and study planning.
    \item A learning-context modeling approach that uses mastery history, practice spacing, task difficulty, and help-seeking signals to guide when and how AI assistance is delivered.
    \item A longitudinal evaluation framework suitable for real-world school deployment (e.g., cluster-randomized or stepped-wedge rollout) that connects learning traces in middle school to medium- and long-term educational outcomes.
    \item A governance strategy for K--12 AI deployment emphasizing privacy-first design, policy-gated assistance, audit logs, and equity analyses to assess differential effects across student subgroups.
\end{itemize}

The remainder of this paper is organized as follows: Section~2 reviews related work on LMS platforms, AI in education, and longitudinal learning studies. Section~3 presents the motivation, conceptualization, and learning-context model guiding the system design. Section~4 describes the system implementation and architecture, including the AI assistance layer and teacher analytics. Section~5 details the deployment and evaluation methodology, including outcomes, fidelity measurement, and validity considerations. Finally, Section~6 provides concluding remarks, limitations, and future directions for multi-year deployments of AI-integrated learning infrastructure in K--12 settings.

\section{Related Work}

This work sits at the intersection of Learning Management Systems (LMSs), Intelligent Tutoring Systems (ITS), and learning analytics, and emerging generative-AI assistance for classrooms. We review prior evidence on (i) what LMSs do well and where they fall short for moment-to-moment instruction, (ii) what ITS research suggests about scalable individualized support, (iii) how feedback, spacing, and help-seeking mechanisms relate to durable learning, (iv) how teacher-facing analytics enable human-in-the-loop orchestration, and (v) why longitudinal evaluation and governance are essential when deploying AI with minors.

\subsection{LMSs, Learning Analytics, and the Instructional Gap}
LMS platforms are widely used to distribute materials, collect submissions, and report grades, improving coordination and record-keeping \cite{DahlstromBrooks2014ECAR}. However, conventional LMS workflows are primarily administrative and often provide limited real-time instructional guidance when students are stuck \cite{DahlstromBrooks2014ECAR}. This gap matters because formative feedback is most effective when timely and actionable; delays can weaken its impact on learning and revision behavior \cite{DahlstromBrooks2014ECAR,HattieTimperley2007}. Learning analytics has aimed to close this gap by transforming trace data (attempts, time-on-task, completion patterns) into signals for students and teachers, yet many deployments still struggle to translate dashboards into actionable classroom interventions at the right moment \cite{DahlstromBrooks2014ECAR,HattieTimperley2007}. In practice, the extent to which these signals translate into timely, in-class support depends heavily on how schools and teachers integrate them into everyday instruction.

\subsection{Intelligent Tutoring Systems and Evidence for Personalized Support}
ITS research provides a strong empirical basis for scalable, individualized learning support. Meta-analyses indicate that tutoring systems can improve learning outcomes compared to business-as-usual instruction, although effects vary by domain, implementation quality, and the comparison condition \cite{VanLehn2011MetaAnalysisTutoring,Ma2014ITS}. A recurring lesson is that benefit depends less on having help than on delivering it well—timed and framed to support productive struggle rather than simply supplying answers  \cite{VanLehn2011MetaAnalysisTutoring,Aleven2003HelpSeeking}. This motivates bounded assistance policies and context-aware hinting that responds to student errors while preserving opportunities to learn \cite{Aleven2003HelpSeeking}.

A complementary line of work tests tutoring technologies in authentic school settings. Large-scale evaluations of deployed tutoring curricula (e.g., Cognitive Tutor Algebra I) highlight both the promise and the practical complexity of adoption, including implementation-fidelity variation and heterogeneous impacts across sites and cohorts \cite{Pane2014CognitiveTutor}. These findings reinforce the need for instrumentation (to measure actual exposure) and study designs that remain robust under typical school constraints \cite{Pane2014CognitiveTutor}.

\subsection{Feedback, Help-Seeking, and Spaced Practice as Mechanisms}
Our system design operationalizes three mechanism families with deep roots in learning sciences and AIED: feedback, self-regulated help-seeking, and spacing. Feedback research highlights that effective feedback answers questions about goals, progress, and next steps, and it is most useful when specific and timely \cite{HattieTimperley2007}. Help-seeking is a key self-regulated learning strategy, but it is also a frequent failure mode: students may avoid help, overuse hints, or seek answers rather than guidance. Prior ITS work has therefore modeled and scaffolded help-seeking behavior to encourage learning-oriented requests and reduce gaming behaviors \cite{Aleven2003HelpSeeking,Karabenick2003HelpSeeking}. Spaced practice research further shows that distributing review over time improves retention and transfer; computational models have been used to schedule practice adaptively based on forgetting dynamics and performance history \cite{PavlikAnderson2008Spacing}. Together, these strands motivate a policy that prioritizes bounded hints, revision prompts, and review scheduling over direct solution delivery.

\subsection{Teacher-in-the-Loop Orchestration and Classroom Dashboards}
A recurring lesson from AIED deployments is that classroom learning is not purely a student--system interaction. Teachers remain responsible for pacing, motivation, differentiation, and responding to misconceptions. As a result, teacher-facing analytics and orchestration tools have become central to making tutoring support effective in real classrooms\cite{Holstein2018Orchestration}. Prior work on classroom orchestration has examined how real-time dashboards can surface student struggle and support timely teacher action while respecting classroom realities \cite{Holstein2018Orchestration}. This motivates our emphasis on (i) dashboards that summarize misconception patterns and persistent struggle, (ii) intervention tools that translate insights into concrete actions (e.g., targeted review sets), and (iii) logging teacher mediation so that downstream analyses can distinguish direct student-facing effects from teacher-amplified effects\cite{Holstein2018Orchestration}.

\subsection{Generative AI in Education and the Need for Guardrails}
Recent large language models (LLMs) introduce new capabilities for natural-language feedback, explanation, and dialogic help, but they also introduce risks that are especially salient for minors: hallucinated explanations, inconsistent pedagogical quality, privacy leakage via over-sharing context, and answer-giving that undermines learning \cite{Kasneci2023ChatGPTGood}. Analyses of LLMs in education emphasize both opportunity (e.g., scalable support and interactive explanation) and challenge (e.g., reliability, alignment with instructional intent, and governance) \cite{Kasneci2023ChatGPTGood}. These concerns motivate architectural separation (a policy gateway) and auditable controls that bound AI behavior by context (practice vs.\ assessment), role (student vs.\ teacher), and safety constraints \cite{Kasneci2023ChatGPTGood}.

\subsection{Longitudinal Effects, Fadeout, and Evaluation Design}
A persistent challenge in education research is that short-term gains do not always translate into durable improvements; many interventions exhibit partial fadeout as students transition across grades, teachers, and curricula \cite{Bailey2020Fadeout}. Syntheses on persistence and fadeout point to several reasons why early gains can weaken over time: changes in instructional environments, skill substitution, and measurement mismatch, and recommend designs that model trajectories instead of relying on a single endline snapshot \cite{Bailey2020Fadeout}. These findings motivate our time-horizon outcome framework and our emphasis on school-feasible study designs (cluster randomization or stepped-wedge rollout), paired with fidelity measurement and linkage to longer-run institutional outcomes when data-sharing agreements permit \cite{Bailey2020Fadeout}.

\section{System Overview}
The proposed platform is a learning management system (LMS) designed for middle school instruction that utilizes Artificial Intelligence for more iterative and responsive feedback, as well as a longitudinal evaluation of its post-secondary effects. Its primary goal is to provide structured, safe, and scalable learning support that mirrors and enhances day-to-day classroom practice while producing credible evidence about longer-run outcomes. The system couples standard LMS capabilities (course organization, assignments, assessments, communication) with bounded AI features for formative feedback, adaptive practice sequencing, and study planning. In order to surface misconceptions and support timely interventions, it includes teacher-facing analytics. The platform is intended for deployment in authentic school settings, so students, teachers, and administrators can use it as part of routine coursework.

\subsection{User Roles and Core Use Cases}
The system supports four main roles:
\begin{itemize}
    \item \textbf{Students:} complete learning activities, receive feedback and hints, track progress, and follow recommended review schedules.
    \item \textbf{Teachers:} assign activities, monitor progress, review class-level misconceptions, and deliver targeted interventions.
    \item \textbf{School administrators:} manage rosters, policies, permissions, and reporting requirements across classes and schools.
    \item \textbf{Parents/Guardians (optional):} view high-level progress summaries and engagement indicators, without access to sensitive details or AI chat logs.
\end{itemize}

A typical student workflow begins with an assigned activity (practice set, quiz, reading, or project), followed by attempts, feedback, and revision. When a student struggles, the AI assistance layer provides bounded support such as hints, step-by-step guidance constraints (where appropriate), and short conceptual explanations aligned with the curriculum and course objectives. A typical teacher workflow begins with authoring or selecting an assignment, reviewing class progress, and using dashboards to identify patterns (widespread errors in fractions or reading comprehension). Teachers can then intervene by assigning targeted reviews, initiating small-group instruction, or providing individual support.

\subsection{System Modules}
The platform is organized into interconnected modules that support instruction, assistance, analytics, and governance:

\paragraph{Learning Delivery and Assessment Module.}
This module provides standard LMS capabilities: course pages, lesson resources, assignments, quizzes, rubrics, submission handling, and gradebook integration. It supports item-level assessment so performance can be interpreted at the level of specific concepts and skills, not only as an overall assignment score.

\paragraph{AI Assistance Module.}
The AI module provides bounded, age-appropriate support when students get stuck. It follows a “help, not answers” rule: the goal is to guide students toward the next step with formative feedback and scaffolding, without simply giving away solutions. Example functions include:
\begin{itemize}
    
    \item \textbf{Concept explanations:} short explanations tied to the current unit, with examples appropriate for middle school learners.
    \item \textbf{Study planning:} spaced review recommendations and reminders based on mastery history and time since last practice.
    \item \textbf{Reflection prompts:} brief prompts that promote metacognition (e.g., identifying what was confusing and what strategy helped).
    \item \textbf{Hinting and feedback:} targeted hints based on the student’s error type, plus feedback that encourages revision.

\end{itemize}
All AI outputs are logged for auditability, can be constrained by policy (e.g., no direct answers on graded items), and can be configured by teachers and administrators.

\paragraph{Learning Context and Personalization Engine.}
This engine is critical for personalization as it maintains student-level learning state used to tailor support and sequencing. It aggregates signals such as mastery by concept, attempt history, time-on-task, spacing intervals, and help-seeking behavior. The engine determines when to recommend review, when to escalate to teacher attention, and what type of scaffolding is most appropriate for the student’s current context.

\paragraph{Teacher Analytics and Intervention Dashboard.}
The teacher dashboard provides actionable summaries at the student and class levels. Key views include:
\begin{itemize}
    \item \textbf{Progress and mastery:} concept-level mastery estimates and growth over time.
    \item \textbf{Misconception trends:} clusters of common wrong answers and error patterns.
    \item \textbf{Engagement and persistence:} attendance in the platform, completion rates, repeated struggle indicators, and overdue work.
    \item \textbf{Intervention tools:} recommended review sets, targeted mini-lessons, and flags for students needing follow-up.
\end{itemize}

\paragraph{Governance, Safety, and Privacy Module.}
Because the platform aims to enhance the learning of minors, governance and privacy are first-class components that are considered in every component of the system and longitudinal study. This module enforces role-based access control, data minimization policies, encryption, and retention rules. It also provides guardrails for AI behavior, including, but not limited to, topic constraints, age-appropriate responses, no solution disclosure on graded work. The system also produces audit logs to support transparency and oversight by educators and researchers.

\subsection{Data Capture and Logging for Longitudinal Analysis}
To support rigorous evaluation during and after usage, the system logs both instructional outcomes and learning processes. Logged data include assignment completions, item-level responses, mastery estimates, revision events, hint usage, AI interaction metadata, and teacher actions (assignments created, feedback provided, interventions applied). For longitudinal linkage, the system uses stable, privacy-preserving identifiers that allow outcomes to be connected across years without exposing personally identifiable information to researchers.

\subsection{Design Goals and Operating Assumptions}
The system is guided by four design goals: (i) \textbf{pedagogical usefulness} through timely, mastery-oriented feedback and how well these help learning; (ii) \textbf{safety and appropriateness} through bounded AI behavior, minor-appropriate responsiveness, and auditability; (iii) \textbf{equity} through monitoring of subgroup effects and avoidance of differential harm; and (iv) \textbf{deployment realism} through workflows that fit classroom constraints and support variable connectivity and device access. These goals inform both the platform’s features and the evaluation approach described in subsequent sections.

\section{System Architecture}
The AI-integrated LMS is implemented as a modular, service-oriented platform that separates instructional workflows from AI assistance, analytics, and governance. This separation supports safe deployment in K--12 settings by enabling strong access control, auditability of AI outputs, and clear boundaries between learning content, student data, and automated support. At a high level, the architecture consists of (i) client applications for students and teachers, (ii) a core LMS backend, (iii) an AI assistance service constrained by policy, (iv) a learning analytics and personalization engine, and (v) secure data stores and monitoring components.

\FloatBarrier

\begin{figure*}[t]
    \centering
    \includegraphics[width=\textwidth]{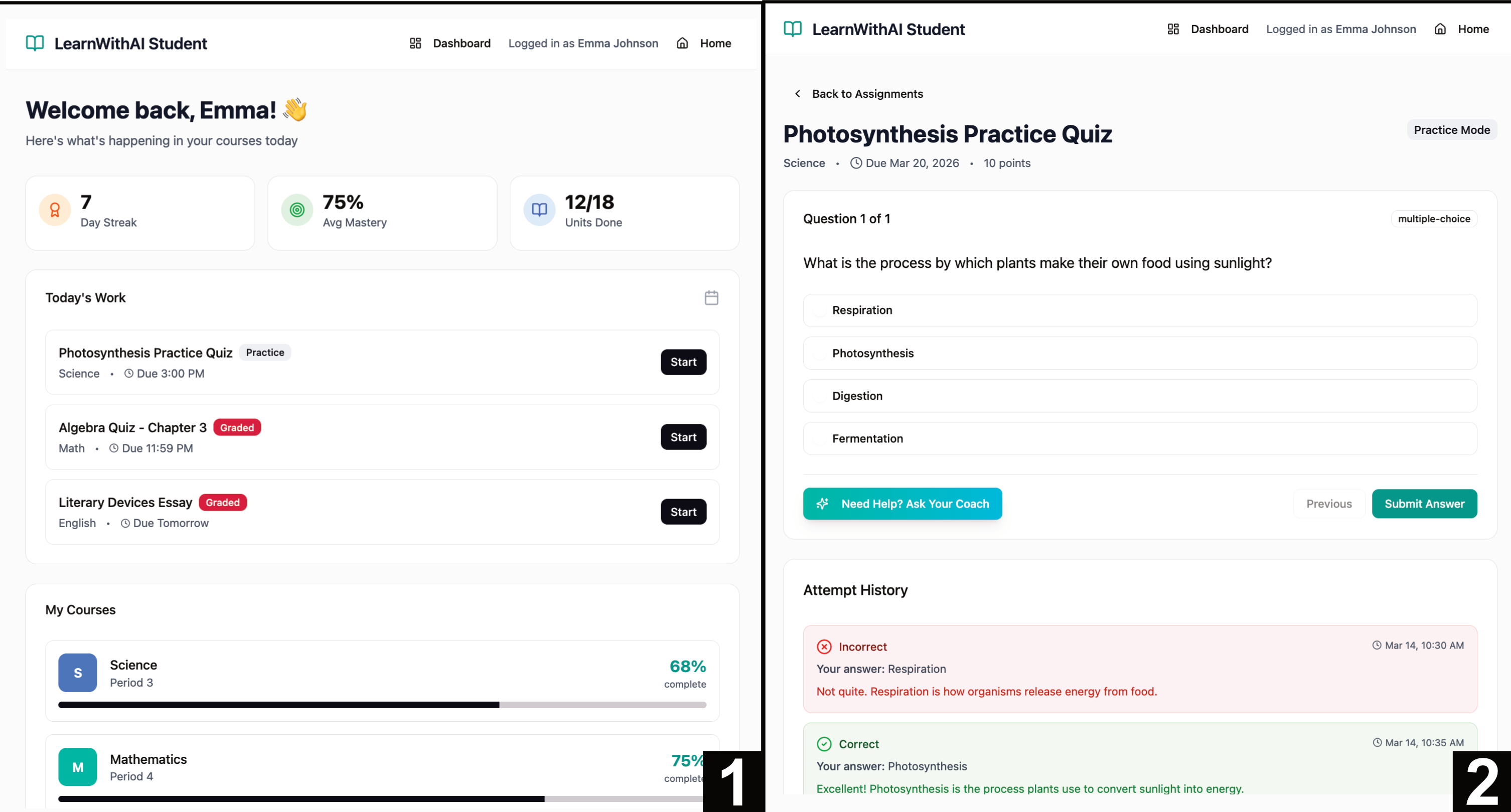}
    \caption{(1) shows the student home dashboard, summarizing streak and mastery, plus “Today’s Work” assignments with one-click start. (2) shows a practice-quiz attempt page, including the question UI, a Practice Mode indicator, and an attempt history panel with correctness feedback.}
    \label{fig:img1}
\end{figure*}

\subsection{High-Level Components}
\paragraph{Client Layer (Student, Teacher, Admin).}
The client layer includes a student learning interface and a teacher/admin portal. Students use the platform to access lessons, complete assignments and quizzes, request hints, and follow study plans, as illustrated by the student home and practice-quiz views in Figure~\ref{fig:img1}. Teachers use the complementary portal to create and schedule assignments, review dashboards, provide feedback, and configure classroom policies (when AI hinting is permitted), with representative teacher dashboards shown in Figure~\ref{fig:img3}. Administrators manage rosters, permissions, and reporting.

\paragraph{AI Assistance Service (Bounded Tutor).}
The AI service provides context-aware formative support, but it is intentionally isolated behind a \textit{policy gateway} so guardrails are enforced before any response reaches a student. When help is requested, the gateway evaluates classroom policy and task context, applying rules such as age-appropriate constraints, content filters, and assessment-specific restrictions (“no direct answers” on graded work). The contrast between practice-mode coaching and graded contexts where help is restricted is shown in Figure~\ref{fig:img2}. The AI receives only the minimal task-relevant context needed to generate support (problem prompt, the student’s current attempt, and a compact learning-state summary). Prompts, responses, and policy decisions are logged to enable auditing, error analysis, and evaluation.

\paragraph{Learning Analytics and Personalization Engine.}
This engine turns raw event logs into actionable learning signals. It estimates concept-level mastery, detects sustained struggle, and recommends next activities, including targeted review and spacing-based practice. It produces the signals needed for personalization (e.g., spaced review schedules, difficulty adjustments) and teacher dashboards (e.g., misconception clusters, at-risk flags), with dashboard examples shown in Figure~\ref{fig:img3}. The engine is also responsible for generating longitudinal metrics, such as growth curves and persistence indicators, used in evaluation.

\paragraph{Data Layer (Secure Storage and Linkage).}
The data layer includes: (i) a transactional store for LMS operations (assignments, grades, rosters), (ii) an event log for fine-grained learning traces (attempts, hints, revisions), and (iii) a research warehouse where de-identified extracts are stored for analysis. Longitudinal linkage is supported via stable, privacy-preserving identifiers managed under strict governance policies. Personally identifiable information (PII) is stored separately from research logs, and access is restricted by role.

\paragraph{Governance, Monitoring, and Audit.}
A governance service enforces role-based access control, consent/assent status, data retention rules, and AI policy configurations. Monitoring includes integrity checks (e.g., missing log events), model safety metrics (e.g., blocked responses, escalation rates), and equity indicators (e.g., differential help usage or outcomes by subgroup). Audit logs capture AI prompts/responses, policy decisions at the gateway, and any teacher overrides.

\subsection{Key Workflows}
\subsubsection{Assignment and Learning Workflow}
Teachers author or select activities aligned with course objectives and curriculum standards and publish them to classes. Students access the activity through the LMS, submit responses, and receive immediate formative feedback. The LMS records each attempt and outcome, while the analytics engine updates mastery estimates and recommends follow-up practice or review based on performance and spacing history.

\subsubsection{AI Help Request Workflow}
When a student requests help, the client sends a help request to the AI policy gateway. The gateway checks classroom policy (e.g., allowed hint types), student consent status, and task context (practice versus graded quiz). It then constructs a minimal context package, calls the AI assistance service, and returns a bounded response (hint, explanation, or question prompt). The system logs the request metadata, the gateway’s decision, and the final output for auditing. If help is not allowed, or the model output violates a constraint, the gateway blocks it and returns a safe fallback (for example, “ask your teacher”) or directs the student to approved resources.

\begin{figure*}[!t]
    \centering
    \includegraphics[width=\textwidth]{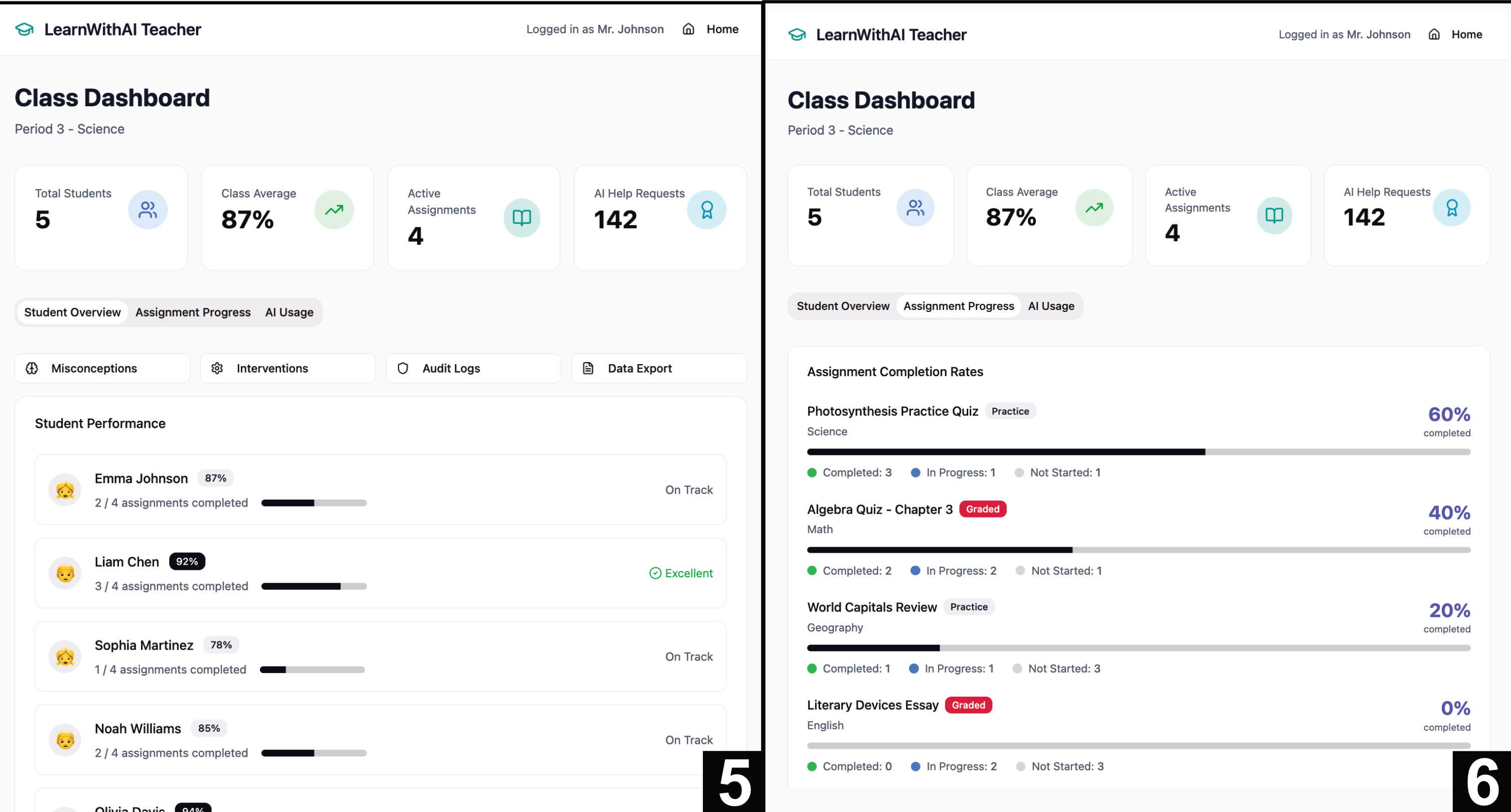}
    \caption{(5) shows the teacher class dashboard (student overview) with per-student completion/progress and status indicators for monitoring. (6) shows the teacher class dashboard (assignment progress) with assignment-level completion rates and breakdowns (completed / in progress / not started).}
    \label{fig:img3}
\end{figure*}

\paragraph{Teacher Monitoring and Intervention Workflow}
Teachers access dashboards that summarize progress, misconceptions, and engagement (Figure~\ref{fig:img3}). When the system detects sustained struggle (e.g., repeated incorrect attempts with high hint usage and low mastery gain), it can flag the student for teacher attention. Teachers can assign targeted review sets, schedule mini-lessons, or provide individualized feedback; misconception clustering and the targeted intervention assignment flow are shown in Figure~\ref{fig:img4}. These intervention actions are logged to support fidelity analysis in the longitudinal study.

\clearpage

\begin{figure*}[!t]
    \centering
    \includegraphics[width=\textwidth]{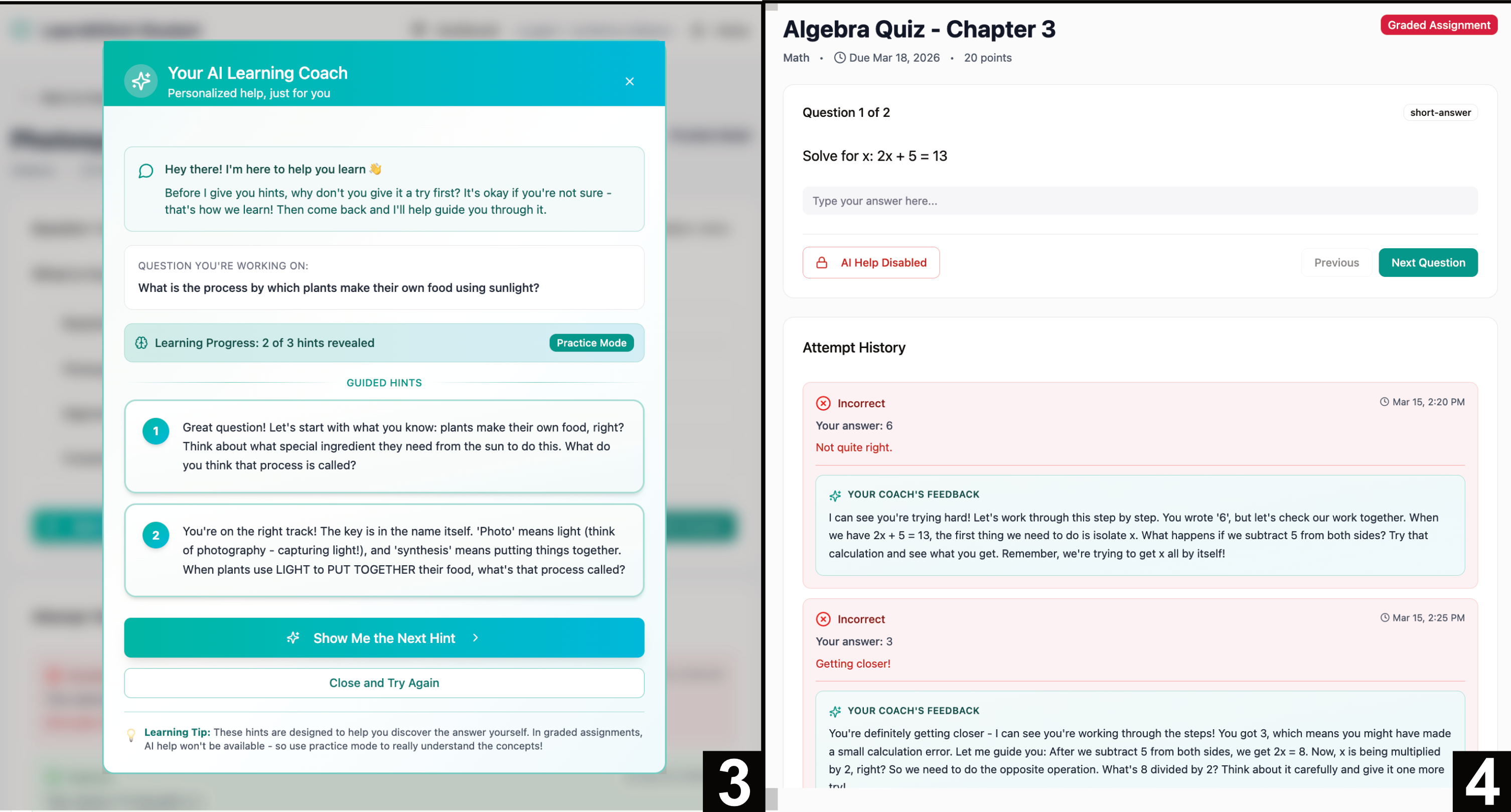}
    \caption{(3) shows the AI Learning Coach modal in practice mode, providing guided, stepwise hints with a controlled “next hint” progression. (4) shows a graded assignment screen where AI help is disabled, while the system still records attempt history and surfaces instructional feedback aligned with the work.}
    \label{fig:img2}
\end{figure*}

\clearpage

\begin{strip}
\centering
\includegraphics[width=\textwidth,height=0.60\textheight,keepaspectratio]{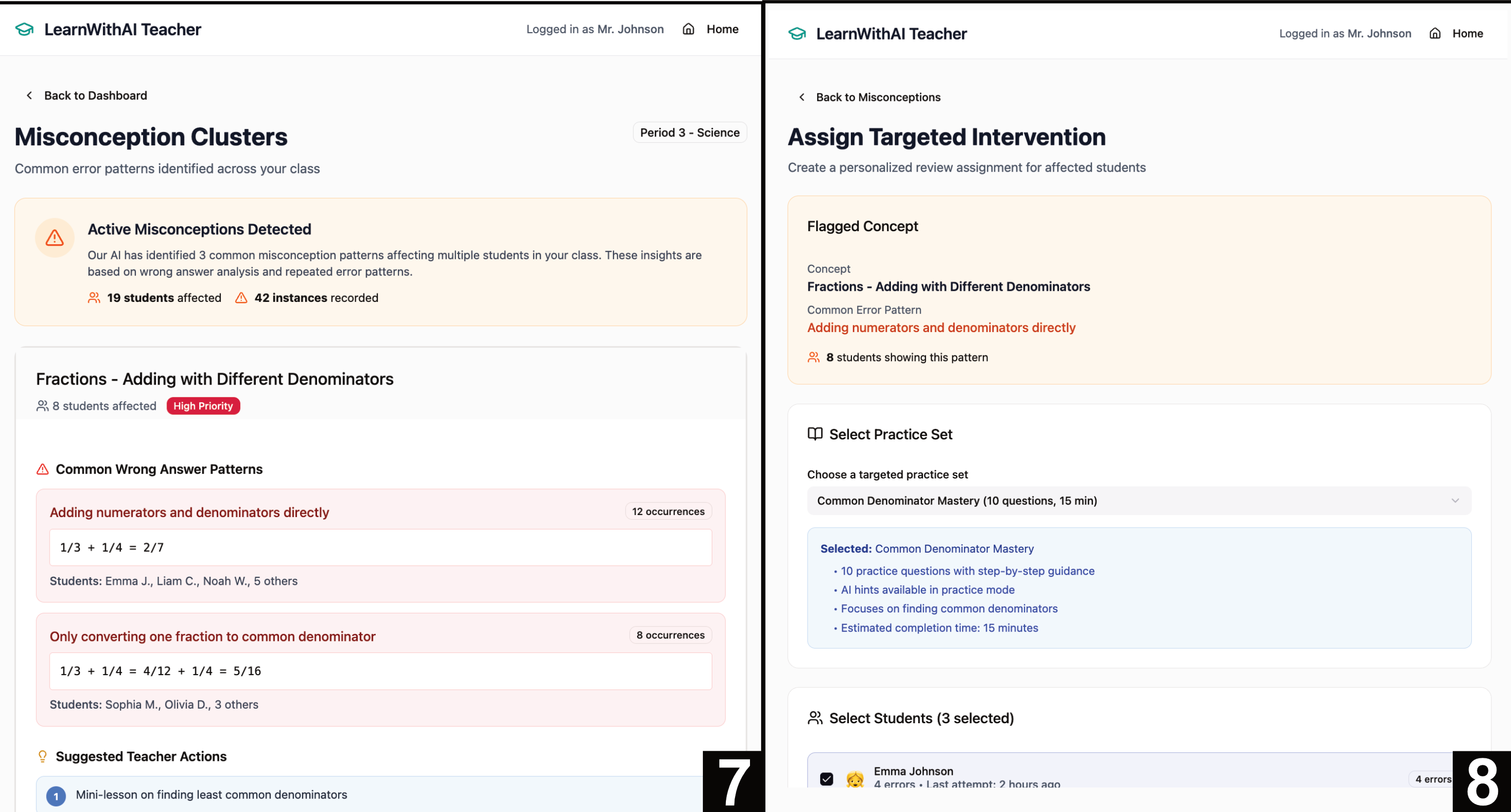}
\captionof{figure}{(7) shows the misconception clustering view, summarizing detected misconceptions, affected students, and representative wrong-answer patterns. (8) shows the targeted intervention assignment flow, where the teacher selects a recommended practice set and assigns it to the affected students for the flagged concept.}
\label{fig:img4}
\end{strip}

\subsubsection{Longitudinal Data Linkage Workflow}
For multi-year evaluation, the platform maintains stable student identifiers within the district/school system while generating de-identified research keys for analysis. Periodic extracts link LMS learning traces to institutional outcomes (course grades, standardized assessments) under approved governance and consent protocols. This linkage enables growth modeling across middle school, high school, and post--high school outcomes while minimizing exposure of PII.

\subsection{Security, Privacy, and Safety Controls}
Because the platform is intended for use by minors and is designed to capture fine-grained learning traces over time, security, privacy, and safety are treated as first-class architectural requirements rather than implementation details. Controls are embedded across the data flow; from what information is collected, to who can access it, to how AI assistance is constrained and reviewed, so that instructional support can be delivered without exposing sensitive student information or permitting unsafe model behavior.

The system follows a \textit{data minimization} principle. The AI assistance layer is intentionally isolated behind a gateway and receives only the context needed to produce bounded support for the current task (e.g., the problem prompt, the student’s latest attempt, and a compact learning-state summary). It does not receive full student profiles, unrelated submissions, or unnecessary identifiers. This reduces the privacy risk surface and limits the consequences of accidental exposure.

Secondly, access to information is governed through strict \textit{role-based access control}. Students, teachers, administrators, and (optionally) parents/guardians are assigned distinct permissions aligned with their instructional responsibilities. For example, teachers may view class dashboards and learning traces needed for intervention, while parent/guardian views can be limited to high-level progress indicators without access to raw AI interaction logs or sensitive text entries. Administrators manage policies, rosters, and compliance settings, but do not require unrestricted access to student-level instructional artifacts unless explicitly authorized.

Sensitive identifiers and linkage keys are protected through \textit{encryption, separation, and retention limits}. Personally identifiable information is stored separately from de-identified research and event logs, and is protected with encryption in transit and at rest. Retention rules are applied so that the system does not store sensitive material longer than necessary for instruction, auditing, and approved evaluation, and so that longitudinal linkage can be achieved without routinely exposing identifiable data to researchers.

In addition to this, all AI assistance is \textit{policy-gated}. Every help request passes through a policy layer that enforces age-appropriate response constraints, task-specific restrictions like avoiding direct answers on graded assessments, and curriculum-aligned behaviors. When a request is disallowed or the model’s output violates a constraint, the gateway blocks the response and provides a safe fallback; this may include prompting the student to re-check a prior step, consult provided resources, or ask the teacher, ensuring that unsafe or instructionally harmful outputs do not reach students.

Finally, the system is designed for \textit{auditability and oversight}. The platform logs AI prompts and responses, policy decisions at the gateway, and relevant metadata (e.g., task type, time, and whether a response was blocked). These logs and records make it possible to review what happened, investigate failures, tune guardrails over time, and confirm that the AI stayed within approved policies. The intent is straightforward: deliver useful support without giving up privacy, human control, or fairness.

\subsection{Deployment Considerations}
The system is designed for realistic school environments with heterogeneous devices and connectivity. The client layer supports low-friction access and can degrade gracefully (e.g., offline caching for reading content, delayed sync for logs). The backend services can be deployed on district-approved infrastructure with separation between operational systems and the research warehouse. These design decisions support scalable deployment while preserving safety, privacy, and the integrity of longitudinal evaluation.

\section{Evaluation Design}
This section describes a longitudinal evaluation plan to estimate the immediate and sustained effects of deploying an AI-integrated LMS in middle school. The primary objective is to determine whether AI-supported learning produces durable gains that persist through high school and influence post--high school pathways, rather than only short-term improvements in engagement or performance. The evaluation, therefore, combines (i) rigorous rollout designs suitable for real school contexts, (ii) multi-level outcome measurement across time, and (iii) analyses that examine mechanisms, heterogeneity, and threats to validity.

\subsection{Study Design and Conditions}
\paragraph{Preferred design: Cluster-randomized trial.}
Where feasible, the study uses cluster randomization at the classroom or school level to reduce contamination, for example, students sharing AI assistance across conditions. Clusters are assigned to one of two conditions:
\begin{itemize}
    \item \textbf{Treatment: AI-integrated LMS} with bounded AI feedback, adaptive practice, study planning, and teacher dashboards.
    \item \textbf{Control: standard LMS} providing the same instructional content and assessment structure but without AI assistance and adaptive sequencing.
\end{itemize}
Randomization occurs prior to the start of the academic term, stratified by school and baseline achievement distributions to improve balance.

\paragraph{Alternative design: Stepped-wedge rollout.}
If randomization is constrained by operational requirements, the study uses a stepped-wedge design in which all participating schools eventually receive the AI-integrated LMS, but adoption is staggered across multiple waves. This supports within-school comparisons over time while increasing feasibility and perceived fairness among stakeholders.

\paragraph{Implementation fidelity.}
Because treatment effects depend on actual usage, fidelity is measured explicitly using system logs (e.g., assignment coverage, weekly active usage, completion rates, and teacher dashboard interaction). Fidelity measures are used both as descriptive indicators and as covariates or moderators in analysis.

\subsection{Participants, Timeline, and Follow-Up}
\paragraph{Participants.}
Participants are middle school students and teachers in collaborating schools. The study targets one or more cohorts starting in grades 6--8, with inclusion defined by enrollment in participating classes and completion of consent/assent procedures. Teachers and administrators participate through platform adoption and brief surveys or interviews.

\paragraph{Timeline.}
The evaluation is organized into three phases:
\begin{enumerate}
    \item \textbf{Baseline (pre-deployment):} collection of prior achievement and demographic/context variables; teacher onboarding; initial surveys.
    \item \textbf{Middle school deployment (1--3 years):} continuous logging of learning traces, periodic benchmark assessments, and end-of-term outcomes.
    \item \textbf{Longitudinal follow-up (high school and post--high school):} annual linkage to institutional outcomes, including course-taking, GPA, standardized assessments, graduation, and postsecondary indicators where available.
\end{enumerate}
Follow-up windows are defined in collaboration with districts to ensure feasible data access and stable identifiers.

\subsection{Outcome Measures}
Our evaluation uses a time-horizon structure because the system is designed to influence both (i) near-term learning performance and behaviors in middle school and (ii) longer-run educational trajectories that unfold as students transition into high school and beyond. Accordingly, we measure outcomes at three levels: short-term outcomes that capture immediate learning and engagement during middle school use, medium-term outcomes that reflect how students progress through high school coursework and requirements, and long-term outcomes that describe post--high school transitions when such data can be obtained under approved agreements. This organization also enables us to test whether any observed benefits persist, attenuate, or grow over time.

\subsubsection{Short-Term Outcomes During Middle School}
In the short term, the primary objective is to determine whether the AI-integrated LMS improves learning and reduces sustained struggle during routine coursework. We therefore measure achievement directly from course-embedded assessments and grades, including unit tests, quizzes, assignment scores, and end-of-term course grades. Because grades can reflect both mastery and classroom-specific practices, we additionally compute concept-level mastery indicators using item-level responses and revisions. These mastery estimates allow us to quantify growth on specific skills (e.g., fractions, algebraic reasoning, reading comprehension) and to examine whether improvements are concentrated in particular concept areas or distributed broadly.

We also treat engagement and persistence as first-class outcomes, because a central claim of the system is that timely feedback and structured support reduce cycles of frustration that often precede disengagement. Using platform logs, we quantify weekly active use, completion rates, time-on-task, and patterns of repeated attempts. Crucially, we distinguish between productive persistence (re-attempts followed by improvement) and unproductive persistence (high attempt counts with stagnant performance). Finally, we collect brief, age-appropriate student surveys to capture constructs that are not fully observable in logs, such as academic self-efficacy, perceived challenge, and study habits. These measures help interpret whether performance changes are accompanied by changes in confidence and learning strategies rather than short-lived test-taking effects.

\subsubsection{Medium-Term Outcomes During High School}
Medium-term outcomes focus on whether early exposure to bounded AI support in middle school translates into improved high school trajectories. We examine academic progression using cumulative GPA patterns, course completion, and credit accumulation, which collectively indicate whether students are successfully moving through graduation requirements. Because one of the most consequential pathways is course-taking, we also track enrollment and persistence in rigorous sequences (for example, mathematics pathways that determine access to advanced STEM coursework). Where comparable assessments exist, we incorporate standardized or benchmark measures to support cross-cohort interpretation and to reduce reliance on local grading variation.

In addition, we track indicators of persistence that often correlate with later completion, such as course failure rates and, where accessible, attendance-related measures (e.g., chronic absenteeism). These outcomes are not treated as direct products of the tutoring features alone; rather, they provide a broader view of whether the system affects the conditions under which learning can accumulate over time.

\subsubsection{Long-Term Outcomes Post--High School}
Long-term outcomes assess whether the intervention influences transitions after graduation, which is ultimately the strongest test of durability. When feasible under data-sharing agreements, we measure postsecondary enrollment (2-year, 4-year, or vocational pathways) and early persistence (e.g., retention to a second term/year or milestone completion). Where available, we also examine remediation or placement indicators, since one hypothesized benefit of stronger middle school foundations is reduced need for remedial coursework. Because post--high school pathways vary by context, we explicitly treat these measures as conditional on feasible linkage and interpret them with careful attention to differences in opportunities and local policy constraints.

\subsection{Mechanisms and Process Analyses}
Outcome comparisons alone cannot explain \emph{why} the system succeeds or fails. To identify mechanisms, we analyze fine-grained learning traces that represent hypothesized pathways from AI support to durable learning. First, we examine self-regulation and pacing by analyzing whether students shift from irregular bursts of work to more consistent practice patterns, including spaced review adherence and reduced last-minute cramming. Second, we examine error-to-revision loops by testing whether feedback; whether it comes as AI-generated hints or teacher guidance surfaced through dashboards, actually leads to substantive revisions and later mastery gains, rather than quick, low-effort resubmissions.

Third, we look at help-seeking quality. Because the system is built to scaffold learning rather than hand out answers, we test whether help requests are followed by stronger subsequent attempts and whether the platform reduces unproductive patterns such as rapid guessing. We also treat teacher mediation as a key moderator by linking dashboard use and intervention actions to student trajectories. This helps separate effects that come directly from student-facing assistance from effects that emerge because the system helps teachers target support more efficiently. Together, these mechanism analyses support interpretation of results and guide iterative refinement of the assistance policies and classroom workflows.

\subsection{Equity and Heterogeneous Effects}
The study evaluates whether effects differ across subgroups and prior achievement levels. Analyses examine heterogeneous treatment effects by baseline performance, school context, and usage patterns. Equity monitoring includes differential access and use (device availability, home connectivity proxies, help usage rates) and differential outcomes. The goal is to detect whether the system narrows, maintains, or widens learning gaps.

\subsection{Analysis Plan}
\paragraph{Primary estimands.}
The primary estimand is the average treatment effect of access to the AI-integrated LMS on academic outcomes (grades, benchmarks, mastery growth) during middle school, with secondary estimands for trajectory outcomes in high school and beyond.

\paragraph{Statistical approach.}
Analyses use hierarchical models appropriate for clustered data (students nested within classes and schools). For longitudinal outcomes, growth modeling is used to estimate differences in trajectories over time. Where stepped-wedge designs are used, models include fixed effects for time and wave to control for secular trends.

\paragraph{Missing data and attrition.}
Given the multi-year horizon, attrition is expected. The study tracks attrition sources (student mobility, dropout, missing records) and applies strategies such as inverse probability weighting, multiple imputation for survey measures, and sensitivity analyses to assess robustness.

\subsection{Threats to Validity and Mitigation}
Evaluating an AI-integrated LMS in real school and learning environments introduces practical threats to validity that can bias estimates of impact if not anticipated in the design. We explicitly account for these threats at both the study-design level (how treatment is assigned and rolled out) and the analysis level (how outcomes are modeled, evaluated, and interpreted), and we treat fidelity and governance logs as core evidence for interpreting results rather than as optional implementation artifacts.

\paragraph{Contamination and spillover.}
A primary risk is contamination between conditions, for example, if students in the control condition gain access to AI assistance through peers, shared accounts, or copied AI-generated content. This is mitigated by assigning treatment at the classroom or school level (rather than individual students) when feasible, reducing opportunities for cross-condition sharing within the same instructional setting. In addition, policy controls at the platform level restrict AI access by roster and condition, and the system logs help-request events to detect anomalous usage patterns. Where contamination is still plausible (e.g., shared devices or informal sharing), we will conduct sensitivity analyses to evaluate how robust findings are to plausible levels of spillover.

\paragraph{Implementation variability and fidelity.}
Treatment effects only show up if students and teachers actually use the system. In practice, adoption will vary; some teachers will assign more activities, some classes will use the platform more consistently, and some students will engage deeply while others barely touch it. That variation creates different “doses” of the intervention and can hide real effects (or make weak effects look stronger than they are). To address this, the study defines and tracks fidelity indicators such as active usage rates, assignment completion coverage, dashboard interaction, and AI help request frequency. These metrics are used to (i) document implementation quality, (ii) support moderator analyses that distinguish whether outcomes differ by usage intensity, and (iii) interpret null results that may reflect low adoption rather than ineffectiveness. Teacher training, onboarding materials, and ongoing support are included to reduce between-classroom variability in how the platform is deployed.

\paragraph{Novelty and short-lived engagement effects.}
New educational technology can produce early engagement spikes that do not translate into sustained learning gains. We therefore evaluate persistence by analyzing outcomes across multiple terms and by testing whether early improvements in mastery, pacing, or completion rates are maintained, amplified, or decay over time. In stepped-wedge designs, time and wave effects are modeled explicitly to separate novelty-related shifts from sustained treatment effects. Longitudinal follow-up measures in high school further reduce the risk of over-interpreting short-term changes.

\paragraph{Measurement drift and changes in assessment regimes.}
Over multi-year studies, outcome measures can drift due to curriculum changes, teacher-created assessments evolving, policy changes, or shifts in test formats. We mitigate this by prioritizing consistent benchmarks when available, anchoring mastery estimates to stable skill taxonomies, and documenting any changes in assessment instruments. When measures change across years, we will use linking strategies (e.g., equating where feasible, or modeling outcomes within-grade and within-year) and will transparently report which outcomes are strictly comparable versus context-dependent.

\paragraph{External confounds and concurrent initiatives.}
Schools often introduce multiple interventions simultaneously (curriculum adoptions, tutoring programs, scheduling changes), and these external factors can confound estimated effects. Randomization helps balance unobserved confounds at baseline when feasible; in stepped-wedge rollouts, time fixed effects help control secular trends. We also collect contextual covariates (e.g., school-level initiatives and policy changes) to support covariate adjustment and to interpret atypical shifts. Where strong concurrent initiatives occur unevenly across sites, we will perform subgroup and sensitivity analyses to assess whether estimated effects are driven by external shocks rather than the platform.

Together, these mitigation strategies aim to preserve internal validity while maintaining deployment realism. Importantly, we treat validity threats as empirically measurable wherever possible (via system logs, fidelity metrics, and contextual documentation), enabling more defensible causal claims and more interpretable findings in real school settings.

\subsection{Ethics, Consent, and Data Governance in Evaluation}
Because the participants are minors, the study requires parental consent and student assent, along with plain-language explanations of what data the system collects and what the AI is allowed to do. Access to raw event logs and AI interaction records is tightly limited to approved personnel, and analyses are run on de-identified extracts under the project’s governance and data-sharing rules. The study also includes continuous safety monitoring of AI outputs, with a clear process for reporting issues and quickly correcting or disabling behaviors that are harmful, inappropriate, or out of policy.

\section{Results and Discussion}
Because the platform described in this paper is \textit{proposed} rather than deployed at scale, the results we report in this section are framed around what the system produces in normal use, what can be observed and logged during operation, and how those signals map directly into the longitudinal evaluation design described later. In other words, the “results” here are the expected outputs of the proposed LMS and the measurable artifacts it is designed to generate when it is used in real classroom workflows, alongside the kinds of findings the longitudinal study is set up to test over multiple years.

As presented in the system architecture, the proposed AI-integrated LMS supports both student-facing learning workflows and teacher-facing monitoring and intervention workflows. On the student side, a learner accesses the course materials and assigned activities through a home dashboard that summarizes progress and provides quick entry into the next tasks (Fig.~\ref{fig:img1}). A typical learning cycle begins when a student starts a practice activity, submits an attempt, and then receives immediate formative feedback that is intended to encourage revision rather than simple answer copying. The practice-quiz interface in Fig.~\ref{fig:img1} illustrates how attempt history is surfaced to the student so the learner can see whether they are improving across tries, which is important for both day-to-day learning and for later analysis of persistence and revision behavior.

A key element of the proposal is that AI support is \textit{policy-gated} and depends on the learning context. In practice mode, the AI Learning Coach is available and is designed to provide bounded, stepwise hints with controlled progression (Fig.~\ref{fig:img2}). The purpose here is not to replace the learning task, but to keep students moving when they are stuck by offering small next-step guidance and short explanations that remain aligned with the curriculum. In contrast, when the student is working on graded items, AI help is disabled by policy (Fig.~\ref{fig:img2}), while the LMS still records attempts and supports teacher feedback and review. This separation is also important for evaluation because it distinguishes learning-time help-seeking from assessment performance, reducing the risk that measured outcomes are driven by direct answer exposure rather than genuine learning.

On the teacher side, the system is designed to support monitoring at both the student and assignment levels. The teacher class dashboard provides an overview of student progress and status indicators for monitoring (Fig.~\ref{fig:img3}). A complementary view summarizes assignment-level completion patterns, including breakdowns such as completed, in progress, and not started (Fig.~\ref{fig:img3}). These views are intended to reduce the friction of day-to-day classroom orchestration by translating activity traces into teacher-readable summaries, making it easier to decide when a class-wide re-teach is needed versus when targeted support is more appropriate.

Beyond monitoring, the proposal includes teacher-facing analytics aimed at turning observations into action. The misconception clustering view summarizes detected misconceptions, the students affected, and representative wrong-answer patterns (Fig.~\ref{fig:img4}). In practice, this is meant to help a teacher move from “students are struggling” to “students are struggling for \emph{this} reason,” using error patterns as a bridge to instruction. Once a misconception is flagged, the targeted intervention assignment flow allows the teacher to select a recommended practice set and assign it to the relevant students for the concept (Fig.~\ref{fig:img4}). This workflow closes the loop from detection to intervention and, importantly, produces traceable teacher actions that can be used later to interpret outcomes (for example, whether learning gains coincide with timely teacher-assigned follow-up work, or whether the system is being used mostly as a passive dashboard).

Since the broader contribution of the paper includes a longitudinal study framework, the proposed system is designed to produce data that supports both short-term and long-term inference. In the short term (during middle school usage), the LMS naturally yields measurable outcomes such as assignment completion, item-level correctness, revision events, and help-seeking traces (request frequency and timing in practice mode), alongside pacing indicators like time-on-task and spacing between practice sessions. These signals can be summarized as changes in mastery over time and contrasted across treatment and control conditions in a cluster-randomized or stepped-wedge rollout. In the medium term (during high school follow-up), the longitudinal linkage plan supports outcomes such as course performance trajectories, course-taking patterns, and credit accumulation, which matter because middle school gains often fade without reinforcement. In the long term (post--high school, where feasible under data-sharing agreements), outcomes such as postsecondary enrollment and early persistence indicators provide a stronger test of whether early, bounded AI-supported learning translates into durable academic benefits rather than short-lived engagement effects.

Finally, this proposed system also makes it possible to discuss results in terms of mechanisms, not only outcomes. For example, if students show improved performance after practice-mode coaching (Fig.~\ref{fig:img2}) and those gains later appear on graded work where AI help is disabled, that pattern is consistent with learning transfer rather than simple answer exposure. Similarly, if teacher dashboard monitoring and targeted interventions (Figs.~\ref{fig:img3} and \ref{fig:img4}) are followed by faster correction of specific misconception clusters, that provides a plausible teacher-in-the-loop explanation for observed gains. These are exactly the kinds of patterns the longitudinal evaluation is designed to detect, separate, and interpret over time, especially in the presence of known threats such as implementation variability, novelty effects, and fadeout across educational transitions.

\section{Conclusion and Future Work}
This paper proposes an AI-integrated learning management system (LMS) intended for middle school deployment and a rigorous longitudinal evaluation of learning outcomes through high school and into post–high school pathways. The central premise is based on the formative period that is middle school, for both foundational academic skills and learning behaviors, and how bounded AI support embedded in everyday coursework can provide timely scaffolding that is difficult to deliver consistently at scale. 

The proposed system combines standard LMS capabilities with policy-gated AI assistance for formative feedback and hinting, adaptive sequencing and spaced review, study planning and reflection prompts, and teacher-facing analytics that surface misconceptions and guide interventions. By separating core LMS functions from the AI assistance layer and embedding governance, auditability, and privacy controls, the architecture supports real-world K–12 deployment with safeguards appropriate for minors.

Beyond the system itself, the paper outlines an evaluation design aimed at a question that often remains unresolved in short-term educational technology research: whether early AI-supported learning yields durable benefits across key educational transitions. The proposed study links fine-grained learning traces (attempts, revisions, help-seeking, and pacing) with long -term institutional outcomes (grades, assessments, course-taking trajectories, and postsecondary indicators where feasible). This combination supports both outcome estimation and mechanism analysis, enabling the study to distinguish short-lived engagement effects from sustained improvements in mastery and self-regulation. The evaluation plan also emphasizes implementation fidelity and equity monitoring, recognizing that differential access, teacher adoption, and subgroup impacts may shape both the magnitude and distribution of effects.

Several directions remain for future work. First, the system will require iterative refinement of guardrails and pedagogical policies that govern AI behavior in different contexts, particularly to keep assistance supportive without becoming answer-giving or fostering over-reliance. Second, future deployments should broaden coverage across subject areas and activity formats, including writing, project-based learning, and collaborative tasks, while preserving auditability and teacher oversight. Third, personalization should be explored carefully and indeptthly, balancing improved adaptivity against risks to privacy and fairness, including whether personalization strategies amplify or reduce existing achievement gaps. Fourth, the longitudinal design can be strengthened by improving data linkage strategies across districts and postsecondary partners, enabling more complete tracking of outcomes beyond high school. Finally, future work should examine adoption constraints in schools, including teacher workload, device access, connectivity, and policy compliance, and develop implementation playbooks that support reproducible deployments.

Overall, this work positions AI not as a replacement for teachers or curricula, but as a bounded support layer within an LMS that can help students practice and iterate effectively, recover from misunderstandings, and build durable learning habits. With careful governance and rigorous longitudinal evaluation, the proposed approach can contribute both a deployable system and empirical evidence on whether AI-integrated learning infrastructure can produce sustained educational benefits.

\end{document}